\documentclass[mathleft]{an}
\usepackage{graphicx}
\usepackage{times}
\def\beq{\begin{equation}}
\def\eeq{\end{equation}}
\def\(({\left (}
\def\)){\right )}
\def\I{{\sc i}}
\def\II{{\sc ii}}
\def\cm1{$\rm cm^{-1}$}
\def\kms{$\rm km\,s^{-1}$}
\def\DE{D\kern-0.75em \raisebox{1.0pt}{=}\ }

\def\Sum{N_{\rm tot}}
\begin{document}
% The following seven commands are intended for editorial usage and should be ignored by
% the author(s).
\Pagespan{1}{}% Document's page range. 
% If second parameter is left empty, the last page is computed automatically.
\Yearpublication{2011}%
\Yearsubmission{2011}%
%\Month{9}%   
%\Volume{332}%  
%\Issue{9/10}% 
% \DOI{10.1002/asna.201111589}% 
\title{The absorption and emission spectrum of the magnetic Herbig Ae star HD 190073%
\thanks{Based on ESO Archival data, from ESO programme076.B-0055(A)
and programme 082.D-0833(A)}}
\author{C.R. Cowley\inst{1}\fnmsep\thanks{Corresponding author:
  {cowley@umich.edu}}
%Example 
%for footnote, note the usage of the \texttt{fnmsep}
%command as separator between institute number and footnote mark} 
\and  S. Hubrig\inst{2}
}
\titlerunning{HD 190073 (V1295 Aql)}
\authorrunning{C.R. Cowley \& S. Hubrig}
\institute{
Department of Astronomy, University of Michigan, Ann Arbor, MI 48109-1042, USA
\and
Leibniz-Institut f{\"u}r Astrophysik Potsdam (AIP), An der Sternwarte 16, 
D-14482 Potsdam, Germany}
%%\and 
%%Downing Street 10, London, UK
%%\and 
%%The second affiliation of the second author}
\received{2011 Sep 16}
%\accepted{2011 Sep 26}
%\publonline{2011}
\keywords{stars: abundances -- stars: individual (HD\,190073) -- stars: pre-main sequence}
%%Editorial notes -- instruction for authors}
\abstract{%
We determine abundances from the absorption spectrum of the 
magnetic Herbig Ae star HD\,190073 (V1295\,Aql).  The observations 
are primarily from HARPS spectra obtained at a single epoch.
We accept arguments that the presence of numerous emission lines does 
not vitiate a classical abundance analysis, though it likely 
reduces the
achievable accuracy.   Most abundances are closely
solar, but several elements show departures of a factor of 
two to three, as an earlier study has also shown.
We present quantitative measurements of more than 60 emission
lines, peak intensities, equivalent widths, and FWHM's. 
The latter range from over 200 \kms (H$\alpha$, He D$_3$)
down to 10--20 \kms (forbidden lines).  Metallic emission
lines have intermediate widths.  
We eschew 
modeling, and content ourselves with a presentation of 
the observations a successful model must explain.
Low-excitation features such as the 
Na \I\ D-lines and [O \I] appear with He \I\ D$_3$, suggesting
proximate regions with widely differing $T_{\rm e}$ and $N_{\rm e}$
as found in the solar chromosphere.  The [O \I] and
[Ca \II] lines show sharp, violet-shifted features.  
Additionally, [Fe \II] lines appear to be weakly present
in emission.}

%%  This article gives instructions for authors of {\it Astronomische Nachrichten} (AN) how
%% to prepare an article according to the current \LaTeX\ class.  The source code of this
%%  paper may be used by the authors of AN as a template.  For further information about the
%% journal, its publisher, its editorial and advisory boards, please look at the World Wide
%%  Web (URL {http:/\slash{}www.aip.de\slash{}AN/}) where this text and accompanying class files
%%  can be obtained from.}

\maketitle

\section{Introduction}
\label{sec:intro}

The intriguing spectrum of the magnetic Herbig Ae star HD\,190073 (V1295\,Aql) has attracted the
attention of classical as well as modern spectroscopists (Merrill 1933;
Swings \& Struve 1940; Catala et al. 2007, henceforth CAT).  
Pogodin, Franco \& Lopes (2005, henceforth P05)
give a detailed description of the spectrum along with a historical
resume of investigations from the 1930's.  We note that the nature of
HD\,190073 as a young, Herbig Ae star became widely recognized some
three decades after Herbig's (1960) seminal paper.  

HD\,190073 was included among the 24 young stars studied for abundances
by Acke \& Waelkens (2004, henceforth, AW).  
In this important paper, the authors made
the bold assumption that abundances could be determined for stars of this
nature using standard techniques-plane parallel, one dimensional models,
in hydrostatic equilibrium.  The models were used to obtain abundances
from absorption lines with equivalent widths less than 150 m\AA.
These assumptions might very well be questioned.  Material is being 
accreted by these young stars, and the infall velocities are thought
to be near free-fall, several hundred km\,s$^{-1}$.  Does this infall
produce shocks and heating of the atmospheres
that could invalidate models that neglect such complications?  

AW nevertheless proceeded.  Although they did not state
this explicitly, the justification for their assumptions is empirical,
and may be found in their results.  Basically, these are the fact that
their approach yields entirely reasonable stellar parameters and 
abundances, including agreement from lines of neutral and ionized
elements.  Stated simply, their assumptions led them to self-consistent
results.   We make these same assumptions in the present work, taking
some comfort in the fact that self consistency is all one ever has in 
science.

While AW's studies were both competent and thorough, better observational
material is currently available, making it possible to use systematically
weaker lines, and to study more elements.  We have also made use of the 
wings of the Balmer lines, not used by AW.  

The lower Balmer lines have central emissions.  In the case of 
H$\alpha$, the emission dominates the feature.  The Balmer lines
and especially H$\alpha$, have been extensively studied 
(e.g. P05; Cuttela \& Ringuelet 1990).

In the present paper we also study the weaker metallic emission lines,
to provide information on the physical conditions where this
emission occurs.  This was discussed by
CAT who were primarily
concerned with the magnetic field of HD 190073. 
They also give a detailed
qualitative description of the metallic emission lines (primarily Ti \II,
Fe \I/\II\ (cf. Sect.~\ref{sec:emiss} and following).

Hubrig et al. (2006, 2009) reported a longitudinal magnetic field of 84$\pm$30 Gauss, up to
104$\pm$19 Gauss, while CAT found a longitudinal
field of 74$\pm$10 Gauss.

\section{Observations}
\label{sec:obs}

\begin{figure}
\includegraphics[width=55mm,height=83mm,angle=270]{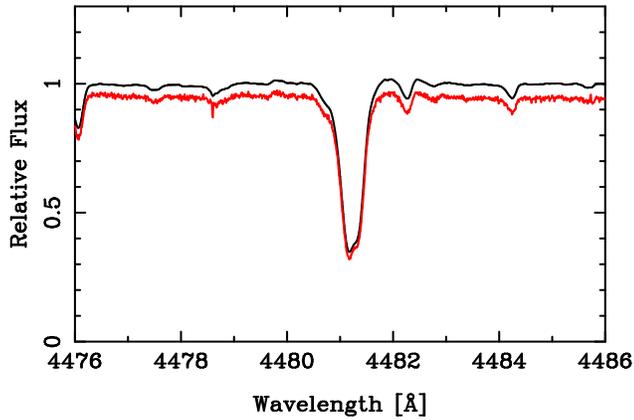}
 \caption{ The HARPS spectrum (ADP.HARPS.2008-11-10T23:43:14.386\_2\_SID\_A)
and averaged spectra in the region of the Mg\,\II\ doublet
$\lambda$4481\,\AA.  The HARPS (gray and red in online version) spectrum has been 
displaced slightly
downward for display purposes.}
\label{fig:line4481}
\end{figure}

We downloaded 8 HARPS spectra from the ESO archive, all obtained on
11 November 2008 within 74 minutes of one another.  These were averaged, binned
to 0.02\,\AA\,
and mildly Fourier filtered.  The resulting spectrum has a signal-to-noise
ratio of 350 or more.  The resolution of HARPS spectra usually cited as 
over 100\,000, is not significantly modified for our purposes on the
averaged spectrum, as Fig.~\ref{fig:line4481} illustrates.

UVES spectra, obtained on 18 September 2005 cover the region from
3044 to 10257~\AA.  They were used for special features (e.g. 
the [Ca \II] lines), but not for abundances.

\section{Reduction}
\label{sec:reduction}

\begin{figure}
\includegraphics[width=54mm,height=83mm,angle=270]{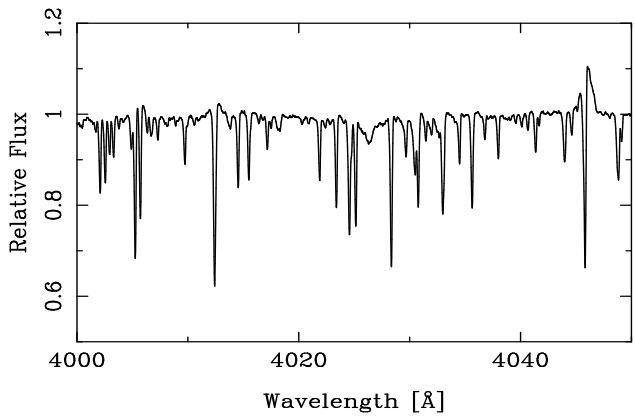}
\caption{The region from $\lambda\lambda$4000 to 4050 \AA\ shows
numerous measurable absorption features.  Note the broad
absorption at $\lambda$4026 \AA, which is He \I.  Fe \I\ 
$\lambda$4045 \AA\ shows strong emission as well as absorption.}
\label{fig:line4050}
\end{figure}

The averaged HARPS spectrum was measured for 1796 lines.  
The UVES spectrum was also measured
for line identifications in the region $\lambda\lambda$3054--3867 \AA.
We measured 760 absorption lines, which were often severely 
affected by emission.  

Many absorption lines, especially weak ones, were not significantly 
affected by emission, and suitable for abundance determinations.
Figure~\ref{fig:line4050} shows a typical region with many relatively
unperturbed absorptions.
Preliminary,
automated identifications were made, and wavelength coincidence
statistics (WCS, Cowley \& Hensberge 1981) were performed.
A few spectra not investigated by AW were analyzed: He \I,
Na \I, Al \I, Si \I, S \I, S \II, Co \I, Mn \I, Mn \II, 
Ni \I, Zn \I, and Sr \II.  We found no exotic elements, such as lanthanides,
or unusual 4d or 5d elements.  

\begin{figure}
\includegraphics[width=54mm,height=83mm,angle=270]{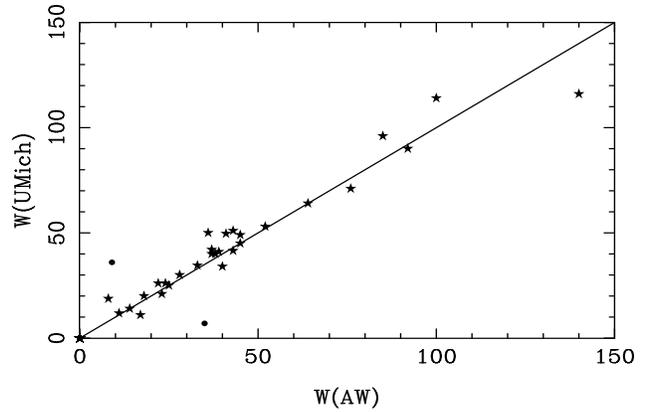}
\caption{ A comparison of equivalent width measurements by AW and
the present study (UMICH).}
\label{fig:pltdif}
\end{figure}

Lines were chosen for equivalent width measurement with the help of
the automated identification list, which lists plausible identifications
within 0.03~\AA\, of any measured feature.  Blends were rejected.  
Usually, we avoided lines with equivalent widths greater than 20 m\AA\,
but in order to compare our measurements with those of AW, we 
included a few stronger lines.  

A comparison of measurements is given
in Fig.~\ref{fig:pltdif}.
Generally, the measurements agree well with one another,
and differences can usually be explained by judgments of where to
draw the continuum when a line is partially in emission, or there
is emission close by.  Differences in the case of one of the solid
circles is surely due to emission, as Ti \II\ $\lambda$4398~\AA\ falls
between two strong emission lines.  The other solid point is for
O \I\, $\lambda$3947 \AA.  This is apparently a misidentification.  
Note that Fig.~\ref{fig:pltdif} is not logarithmic.

\section{The model atmosphere and abundance methods}
\label{sec:model}

The methods used to obtain abundances from the equivalent widths,
including model atmosphere construction are explained in some detail
in two previous papers (Cowley et al. 2010a,b).  Briefly, the 
$T(\tau_{5000})$ from Atlas 9 (Kurucz 1993) as implemented
by Sbordone et al. (2004) was used with Michigan software to
product depth-dependent models.  The effective temperature and
gravity were selected from ionization and excitation equilibrium
as well as fits to the wings of H$\beta$--H$\delta$.  

We have adopted a somewhat lower temperature than used by AW,
8750 K, and $\log g = 3.0$.  The former used 9250 K, and 
$\log g=3.5$.  We also used a lower microturbulence, 
2 km\,s$^{-1}$, compared to AW's 3 km\,s$^{-1}$, but this is not
important for most of our weaker lines.  Oscillator strengths
were taken from the modern literature when possible, or from
compilations by 
NIST (Ralchenko 2010, preferred) or VALD (Kupka et al. 1999).  
Default damping
constants were used as in the studies cited, but they are
unimportant for weak lines.

\section{Abundances}
\begin{table}
\caption{Abundances in HD\,190073 from the current study and AW.}
\label{tab:abund}
\begin{tabular}{l c c r r r} \hline\noalign{\smallskip}
Ion   &$\log({\rm El}/{\Sum})$ & sd  & $N$  &  Sun  &  AW  \\[1.5pt]  \hline\noalign{\smallskip}
      He \I & --1.15           &0.38 & 2  & --1.11 &      \\
      C \I  & --3.40           &0.23 &36  & --3.61 & --3.55  \\
{\bf N \I}* & --3.50           &0.38 & 9  & --4.21 & --3.40 \\
      O \I  & --3.29           &0.10 &12  & --3.35 & --3.38\\
{\bf Na \I} & --5.25           &0.24 & 5  & --5.80 &       \\
      Mg \I & --4.29           &0.23 & 3  & --4.44 & --4.52  \\
      Mg \II& --4.54           &0.16 & 8  & --4.44 &       \\
{\bf Al \I}*& --6.07           &     & 1  & --5.59 &  --6.01\\
      Si \I & --4.43           &0.36 & 7  & --4.53 & --4.41 \\
      Si \II& --4.61           &0.13 &10  & --4.53 &       \\
      S \I  & --4.62           &0.06 & 3  & --4.92 &      \\
      S \II & --4.40           &0.45 & 6  &       &      \\
      Ca \I & --5.78           &0.11 & 2  & --5.70 & --5.41\\
      Ca \II& --5.63           &0.19 & 6  & --5.70 &       \\
{\bf Sc \II}*& --9.16           &0.13 & 10 & --8.89 & --9.00\\
      Ti \II& --7.18           &0.19 & 32 & --7.09 & --7.07 \\
      V \II & --8.07           &0.11 & 14 & --8.11  &--7.93   \\
      Cr \I & --6.54           &0.12 &  6 & --6.40 & --6.35  \\
      Cr \II& --6.37           &0.17 & 22 & --6.40 &         \\
      Mn \I & --6.60           &0.28 &  3 & --6.61 &        \\
      Mn \II& --6.53           &0.24 & 13 & --6.61 &        \\
      Fe \I & --4.54           &0.15 &182 & --4.54 & --4.53  \\
      Fe \II& --4.54           &0.21 &145 & --4.54 &        \\
      Co \I & --7.13           &     &  1 & --7.05 &         \\
      Ni \I & --5.86           &0.15 & 18 & --5.82 & --5.73   \\
      Ni \II& --5.68           &0.23 &  5 & --5.82 &         \\
      Zn \I & --7.48           &0.08 &  2 & --7.48 &       \\
      Sr \II& --8.53           &0.54 &  2 & --9.17 &        \\
      Y \II &--10.05\enspace   &0.19 &  8 & --9.83 & --9.79  \\
      Zr \II&--9.43            &0.15 & 13 & --9.46 & --9.12  \\
      Ba \II&--9.88            &0.12 & 3  & --9.86 & --9.72  \\[1.5pt]  \hline
\end{tabular}
\end{table}

The AW abundances have been converted from differential values, using
Anders \& Grevesse (1989) abundances, which AG adopted.  Our
abundances (see Table \ref{tab:abund}) refer to the Asplund et al. (2009) scale.
The case is not strong that any of these abundances differ
significantly from the solar abundance.  Nevertheless, we have highlighted
in bold face some elements that deserve additional attention.  Nitrogen,
in particular, deserves attention, as it has been found in excess in
the Herbig Ae star HD 101412 (Cowley et al. 2010a). 
Asterisks mark cases where AW and the present work agree on
possibly significant departures from solar abundances.   NLTE effects 
could also be responsible for some non-solar abundances (Kamp et al. 2001).

\subsection{Neutral helium}
\label{sec:he1}

\begin{figure}
\includegraphics[width=54mm,height=83mm,angle=270]{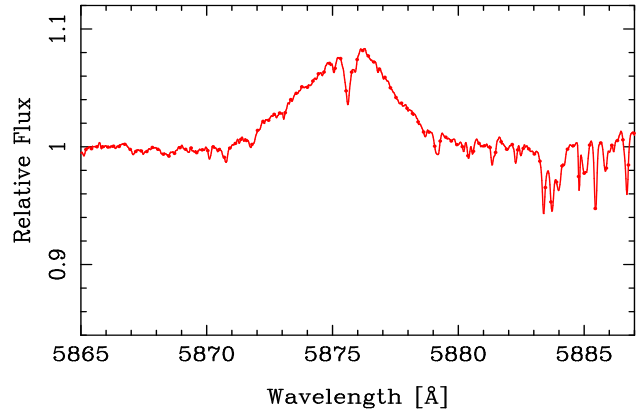}
\caption{ He\,\I\ D$_3$ line in the spectrum of HD\,190073. The central absorption is arguably 
photospheric, and agrees in shape and strength with a
calculated absorption profile.}
\label{fig:d3}
\end{figure}

The helium abundance is from $\lambda\lambda$4026 
(see Fig.~\ref{fig:line4050}) and 4713 \AA.  Both lines
are weak, but in excellent agreement with one another.  However 
$\lambda$4471 \AA\ was also found in absorption, and analyzed.  The
value of $\log({\rm He}/\Sum)$ from this line was found to be $-$1.53, 
some 0.4 dex below the mean of the other two lines.  We have chosen
to disregard this value, as possibly weakened by partial emission.
Should it be included in the average, we find 
$\log({\rm He}/\Sum) = -1.29\pm 0.21$\,(sd), still solar, within the errors. 
The D$_3$ line ($\lambda$5876 \AA) of He \I\, is in emission, and 
included in Table~\ref{tab:lindat}.  It was observed at numerous phases by 
P05, whose observations (see their
Fig. 3) do not show the central reversal clearly seen in our
Fig.~\ref{fig:d3}.  This feature was measured at $\lambda^*$5875.60,
virtually unshifted from the expected photospheric position 
$\lambda$5875.64.  Moreover, an LTE synthesis using Voigt profiles 
and an assumed solar abundance matches the observed absorption
in shape and strength.  Given the likelihood of NLTE, it is
unclear how seriously to take this result.  Nevertheless, the
D$_3$ absorption is consistent with photospheric absorption,
and a solar abundance.

The D$_3$ emission is remarkably similar in morphology to the 
metallic emissions (with unshifted absorptions), though it is 
much broader.  It resembles the P05 illustrations at phases `a'
or `e', with a maximum shifted somewhat to the red, and a longer
violet tail.

\newpage
\section{The emission spectrum}
\label{sec:emiss}

A second focus of the current paper is the emission
spectrum, in particular, permitted 
and forbidden metallic lines.

\begin{figure}
\includegraphics[width=54mm,height=83mm,angle=270]{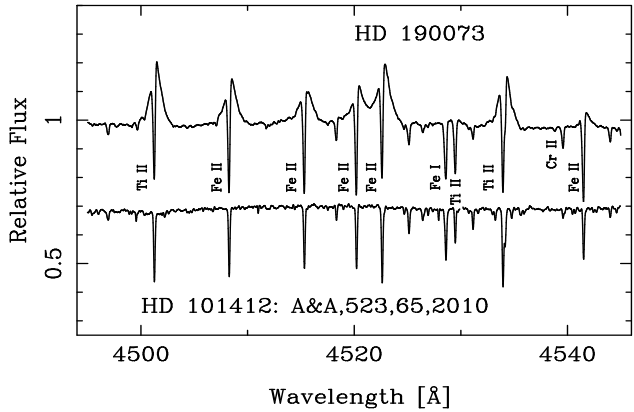}
\caption{The emission/absorption, metallic-line spectrum of 
two Herbig Ae stars, contrasted.  Note the proclivity of the
intrinsically stronger lines to be in emission in HD 190073.}
\label{fig:p2dat}
\end{figure}

%%Much of the previous work has been devoted to 
%%the Balmer lines.
%%so we concentrate on the permitted metallic
%%emission lines and forbidden lines of oxygen and calcium.
%%numerous metallic emission lines, primarily of Fe \I\, and \II,
%%Ti \II, and Cr \II, as well as the Na D-lines.
%%These lines are considered independently of the 
%%stronger emission in the low
%%Balmer members, 

Previous studies (P05, CAT)
provide detailed descriptions of
numerous metallic emission lines, primarily of Fe \I\, and \II,
Ti \II, and Cr \II, as well as the Na D-lines, and have
illustrations similar to the upper spectrum of
Fig.~\ref{fig:p2dat}.  As the P05 work has
a temporal dimension lacking in the present study,
we briefly summarize their findings.  The emissions show mild 
temporal variations both in strengths and widths.  The profiles
are somewhat asymmetric, with their peak intensities generally
very slightly red shifted with respect to the photospheric absorption
spectrum.  The widths of the features are significantly larger 
than these shifts, and a considerable fraction of the emission
is shifted to the blue.  

\begin{figure}
\includegraphics[width=54mm,height=83mm,angle=270]{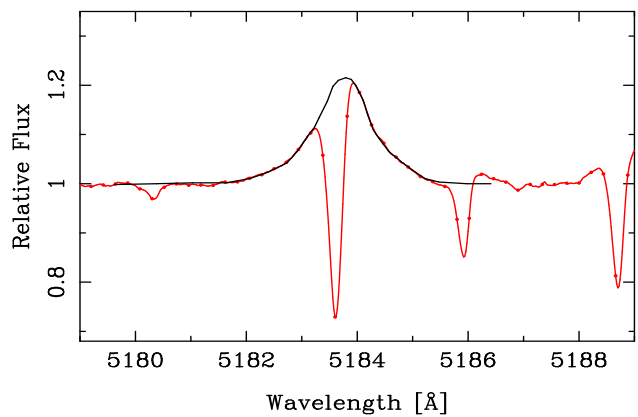}
 \caption{ Segment fits (black) to the emission from Mg\,\I-2 
$\lambda$5183.60 \AA, the strongest of the Mg\,\I\,b lines.
Observations: gray (red online) with dots.
The portion of the fitted curve near the central absorption
is a by-eye estimate of the missing part of the profile.
The maximum designated $I^0$, and
other properties of similar interpolations are
given in Table~\ref{tab:lindat}.}
\label{fig:5183}
\end{figure}

The P05 observations are all in good agreement with the current 
findings, which we take to be a representative sample.
Table~\ref{tab:lindat} gives measurements of the peak intensities,
equivalent widths, and FWHM for the strong relatively unblended
emission lines as they appear
on our averaged HARPS spectra.  Multiplet numbers follow the
spectrum designation.  The intensities are in units of the
continuum, and the equivalent widths are the areas of the 
emissions above the continuum, which is assumed to have unit 
intensity.  The measurements are from segments fitted by eye
to the emission lines, as illustrated in Fig.~\ref{fig:5183}.

All lines in the table had 
central absorptions.  In measuring the $I^0$-values, an attempt
was made to interpolate over this absorption, so the 
$I^0$-value is a few per cent higher than the maximum of
the  emission.  The accuracy of the measurements
vary.  Repeated measurements show consistency for
%I^0$ of a few per cent.  Equivalent widths and
FWHM are generally within 10\,\%.  Underlying emission
from other lines is the prime cause of the uncertainty.

\begin{table}
%\footnotesize
\caption{Maximum intensity measurements, equivalent 
widths, and full widths at half maximum for selected
emission lines.}
\label{tab:lindat}
\footnotesize
\tabcolsep=7pt
\begin{tabular}{c  l  c c  c r} \hline\noalign{\smallskip}
$\lambda$  & Ion/Mult.    & $I^0$ &  $W$[\AA] &  FWHM      & FWHM  \\ 
           &         &       &           &  [\AA]     & [km\,s$^{-1}$\rlap{]} \\[1.5pt] \hline\noalign{\smallskip}
4045.81	& Fe  I-43   &	1.11&	0.130&	1.02&75.6  \\
4063.59	& Fe  I-43   &	1.08&	0.065&	0.83&61.2  \\
4071.74	& Fe  I-43   &	1.11&	0.113&	0.82&60.4  \\
4077.71	& Sr  II-1   &	1.12&	0.109&	0.87&64.0  \\
4143.87	& Fe  I-43   &	1.05&	0.032&	0.61&44.1  \\
4163.65	& Ti  II-105 &	1.05&	0.056&	0.93&67.0  \\
4173.46	& Fe  I-27   &	1.12&	0.095&	0.80&57.5  \\
4178.86	& Fe  II-28  &	1.12&	0.180&	1.26&90.4  \\
4215.52	& Sr  II-1   &	1.06&	0.472&	0.81&57.6  \\
4233.17	& Fe  II-27  &	1.27&	0.402&	1.29&91.4  \\
4246.82	& Sc  II-7   &	1.07&	0.062&	0.82&57.9  \\
4271.76	& Fe  I-42   &	1.06&	0.620&	0.82&57.5  \\
4290.22	& Ti  II-42  &	1.12&	0.112&	0.87&60.8  \\
4294.10	& Ti  II-20  &	1.13&	0.119&	0.82&57.2  \\
4300.05	& Ti  II-41  &	1.23&	0.308&	1.06&73.9  \\
4307.86	& Ti  II-41  &	1.15&	0.179&	1.04&72.4  \\
4351.77	&\llap{*}Fe  I-27   &	1.28&	0.420&	1.31&90.2  \\
4383.55	& Fe  I-41   &	1.14&	0.119&	0.83&56.8  \\
4404.75	& Fe  I-41   &	1.08&	0.077&	0.86&58.5  \\
4443.79	& Ti  II-19  &	1.24&	0.311&	1.06&71.5  \\
4450.48	& Ti  II-19  &	1.06&	0.075&	1.04&70.1  \\
4468.51	& Ti  II-31  &	1.25&	0.281&	0.89&59.7  \\
4491.41	&\llap{*}Fe  II-37  &	1.12&	0.155&	1.15&76.8  \\
4501.27	& Ti  II-31  &	1.24&	0.319&	1.03&68.6  \\
4508.29	& Fe  II-38  &	1.17&	0.198&	1.04&69.2  \\
4515.34	& Fe  II-37  &	1.12&	0.176&	1.25&83.0  \\
4533.97	& Ti  II-50  &	1.19&	0.261&	1.18&78.0  \\
4541.52	& Fe  II-38  &	1.06&	0.048&	0.80&52.8  \\
4558.65	& Cr  II-44  &	1.14&	0.186&	1.17&76.9  \\
4563.76	& Ti  II-50  &	1.16&	0.183&	0.97&63.8  \\
4571.97	& Ti  II-82  &	1.26&	0.332&	1.07&70.2  \\
4576.34	& Fe  II-38  &	1.07&	0.081&	0.97&63.5  \\
4588.20	& Cr  II-44  &	1.09&	0.105&	0.96&62.7  \\
4618.80	& Cr  II-44  &	1.11&	1.350&	1.07&69.4  \\
4629.34	& Fe  II-37  &	1.17&	0.211&	1.00&64.8  \\
4634.07	& Cr  II-44  &	1.04&	0.043&	0.87&56.3  \\
4731.45	& Fe  II-43  &	1.05&	0.058&	0.93&58.9  \\
4805.09	& Ti  II92   &	1.04&	0.332&	0.86&53.7  \\
4824.13	& Cr  II-30  &	1.08&	0.953&	1.06&65.9  \\
4923.92	& Fe  II-42  &	1.65&	1.340&	1.70&103.5  \\
4957.60	& Fe  I-318  &	1.74&	1.590&	1.82&110.1  \\
5169.03	& Fe  II-42  &	1.77&	1.890&	2.14&124.1  \\
5183.60	& Mg  I-2    &	1.23&	0.303&	1.10&63.6  \\
5197.58	& Fe  II-49  &	1.23&	0.231&	1.23&70.9  \\
5234.63	& Fe  II49   &	1.22&	0.437&	1.22&69.9  \\
5264.81	& Fe  II-48  &	1.04&	0.030&	0.66&37.6  \\
5284.11	& Fe  II-41  &	1.08&	0.117&	1.20&68.1  \\
5362.87	& Fe  II-48  &	1.15&	0.216&	1.22&68.2  \\
5534.85	& Fe  II-55  &	1.10&	0.140&	1.10&59.6  \\
5875.64	& He I-4     &	1.09&	0.390&	4.17&212.7  \\
5889.95	& Na  I-D$_2$&	1.76&	1.380&	1.38&70.2  \\
5895.92	& Na  I-D$_1$&	1.70&	1.130&	1.36&69.2  \\
5991.38	& Fe  II-55p &	1.03&	0.043&	1.38&69.1  \\
6238.39	& Fe  II-74  &	1.07&	0.078&	1.14&54.8  \\
6247.56	& Fe  II-74  &	1.13&	0.217&	1.46&70.1  \\
6347.11	& Si  II-2   &	1.11&	0.325&	2.50&118.1  \\
6371.37	& Si  II-2   &	1.08&	0.228&	2.86&134.6  \\
6416.92	& Fe  II-74  &	1.06&	0.098&	1.42&65.7  \\
6432.68	& Fe  II-40  &	1.05&	0.073&	1.16&54.1  \\
6562.82	& H I	       & 6.82&	32.20\enspace\enspace&	5.34&243.9  \\
5158.78 &\llap{[}Fe II-19F] & 1.04& 0.012&\enspace 0.282&16.4 \\
5577.35	&\llap{[}O I-3F]    &	1.01&	0.004&	0.44&23.4  \\
6300.30	&\llap{[}O I-1F]    &	1.11&	0.056&	0.41&19.3  \\
6363.78 &\llap{[}O I-1F]    & 1.04& 0.019& 0.42&19.7  \\
7291.47	&\llap{[}Ca II-1F]  &	1.11&	0.046&	0.43&17.8  \\
7323.89	&\llap{[}Ca II-1F]  &	1.10&	0.033&	0.28&11.3  \\[1.5pt] \hline
%%---------------------------------------------------------
\end{tabular}
\end{table} 

\subsection{Resum\'{e}:  the permitted emissions}

We summarize salient properties of the permitted emission lines:
\begin{itemize}
\item[--] The centers of gravity are shifted by ca. 5 \kms\, to the
red.
\item[--] The profiles are somewhat asymmetric, with a longer violet
than red tail.
\item[--] The central absorption wavelengths are photospheric
within the errors of measurement.  That is, the radial velocities
of the weaker photospheric absorptions agree with those of the central
absorption components of the emission lines.
\item[--] The emission lines are the intrinsically strongest lines.
Weaker lines show weaker emission, until, for the weakest lines,
the observed features are all in absorption.
\end{itemize}
CAT suggested the emissions arose in conditions similar to those
of a photosphere.  They suggested
a heated region 
with densities and temperatures in the range of 
$10^{13}$--$10^{14}$ cm$^{-3}$ and $15\,000$--$20\,000$K.  

Their value of  65 km\,s$^{-1}$ as a typical FWHM for the emissions
agrees well with our measurements (Table~\ref{tab:lindat}).
The origin of this velocity, however, is not readily apparent.
They speculate that these
velocities are due to a supersonic turbulence.  Such ``turbulence''
might arise from the roil of accreting material settling onto the 
photosphere.  
\subsection{Forbidden lines\label{sec:forbidln}}

\subsubsection{The [O \I] lines\label{sec:ForO1}}

\begin{figure}
\includegraphics[width=55mm,height=83mm,angle=270]{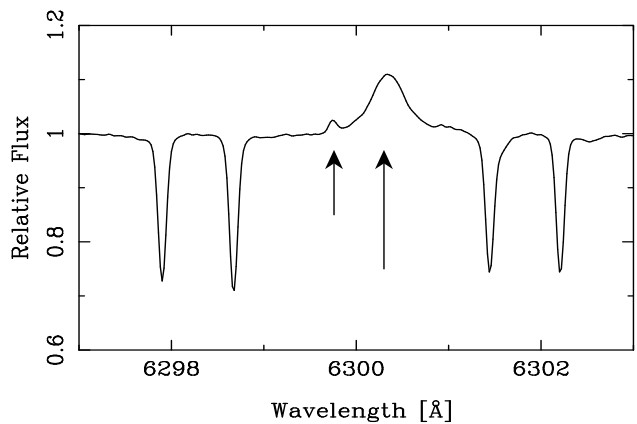}
 \caption{The nebular ($\rm ^3P_2$\,--\,$^1\rm D_2$)
[O \I] transition $\lambda$6300 \AA.  The long arrow points to
the laboratory position at 6300.30~\AA.  The stellar feature
seems slightly red shifted.  The short
arrow points to a narrow, blue-shifted
satellite feature that is also seen at
the same displacement in [O \I] $\lambda\lambda$6363 and 5577 \AA.
The strong absorption lines are atmospheric.}
\label{fig:6300}
\end{figure}

Both $\lambda\lambda$6300 and 6363 [O \I] are present as well
as the auroral transition $\lambda$5577 \AA.  In addition to what
we shall call the main features, all three lines show faint,
sharp, ``satellite'' components shifted to the violet by 
ca. 25\,\kms.  This structure is illustrated in 
Figs.~\ref{fig:6300} and ~\ref{fig:5577d}.  
The main [O \I] features are roughly
one third the width of the typical permitted metallic-line
features, but their peak intensities have comparably small 
red shifts of ca. 5 \kms.  It is plausible to assume the 
[O \I] arises in a region further from the star, and
therefore of lower density than the gas giving rise to the
permitted metallic lines.

The satellite emissions may arise in a polar stream, if we assume
the system is viewed pole on.  The velocity, however, is not high.
The satellite features of all three [O \I] lines is ca. $-25$\, \kms.

\subsubsection{The [Ca \II-F1] doublet}
\label{sec:forbidca}

\begin{figure}
\includegraphics[width=55mm,height=83mm,angle=270]{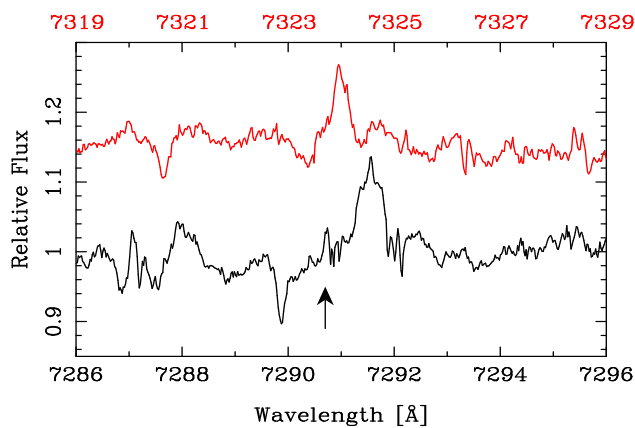}
\caption{The [Ca \II] lines $\lambda\lambda$7291 (black)
 and 7324 \AA\ (gray, red in online version) 
from multiplet 1-F.  Wavelengths for $\lambda$7324 \AA\ are at
the top abscissa. The arrow indicates a possible narrow component
of $\lambda$7291 \AA\ corresponding to those seen in [O \I].}
\label{fig:forca2}
\end{figure}

Hamann (1994) has noted the presence of [Ca \II] in a number
of young stars, including the Herbig Ae V380\,Ori.  We are
not aware that the lines have been previously noted in HD\,190073.
They are also seen in supernovae (Kirshner \& Kwan 1975) and 
extragalactic spectra.
(Donahue \& Voit 1993).  Merrill
(1943) reported [Ca II] lines in emission in the peculiar
hydrogen-poor binary $\upsilon$ß,Sgr (see also Greenstein \&
Merrill 1946).

We find definite emissions at the positions of the forbidden
$\rm ^2S_{1/2}$\,--\,$ ^2\rm D_{3/2,5/2}$ transitions.  
The air wavelengths, determined
from the energy levels, are 7291.47 and 7323.89~\AA.  These
features were identified on a the UVES spectrogram. 
The spectrum (Fig.~\ref{fig:forca2})
is too noisy or blended to show the 
presence of satellite features of $\lambda$7324 \AA, but 
it might be present for $\lambda$7291 \AA.
Measurements of the main
features are included in Table~\ref{tab:lindat}.  The maxima
are shifted to the red by 2 to 4 \kms, in general agreement 
with the [O\,\I] and metallic lines.  The FWHM agree with those
of other forbidden lines.

\subsubsection{[Fe \II]\label{sec:fe2forbid}}
\begin{figure}
\includegraphics[width=55mm,height=83mm,angle=270]{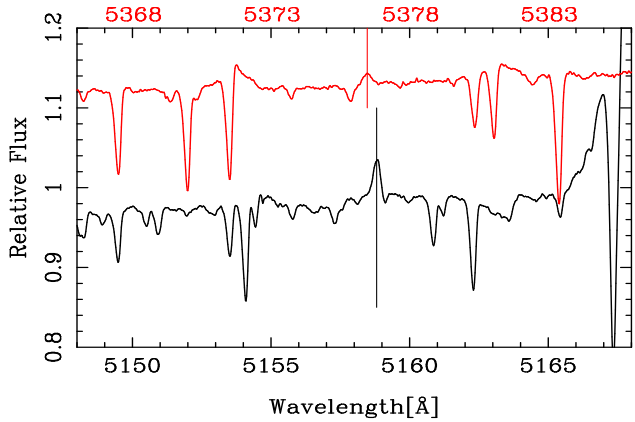}
 \caption{ Forbidden Fe \II\ lines from multiplet 19-F.  The 
wavelength scale for the weaker of the two lines, 
$\lambda$5376 \AA\ (gray, red in online version) is at the top of 
the figure.  Vertical lines mark the positions of wavelengths
derived from the atomic energy levels.}
\label{fig:5158}
\end{figure}

Several workers have discussed [Fe II] emission lines in 
Herbig Ae/Be stars (Finkenzeller 1985; Donati et al. 1997).
We find a definite, sharp emission feature  with a maximum
measured at $\lambda^*$5158.84 \AA.  This wavelength is close to
that of [Fe \II-19F], $\lambda$5158.78 \AA.  (Laboratory positions
for [Fe~\II] are from Fuhr \& Wiese (2006) rather than the
RMT).  Another line from this multiplet is weakly present
(Fig.~\ref{fig:5158}).  Both features are seen on the
unaveraged HARPS spectra as well as UVES spectra taken
some 3 years previously (see Sect.~\ref{sec:obs}).  Several
other lines in Multiplet 19-F are arguably present
($\lambda\lambda$5261, 5296, 5072 \AA), other lines are masked 
by blends or in a HARPS order gap.  The line $\lambda$5158 \AA\
is entered in Table~\ref{tab:lindat}.  It has a FWHM
comparable to that of the other forbidden lines.
     
We found no other [Fe \II] features that could be said to be
unambiguously present.  The well-known [Fe \II] line 
$\lambda$4244 \AA\ is at best, marginally present.

\newpage

\subsubsection{Physical conditions from the forbidden lines}
``Critical electron densities'' are obtained from observed 
forbidden transitions by equating the Einstein spontaneous 
decay rate to the collisional deexcitation rate.  Typical 
values are given by Draine (2011) for $T_{\rm e} = 10\,000$K.  For
the [O\,\I]\, $\rm ^1D_2$-level (the upper level of 
$\lambda\lambda$6300 and 6363 \AA), the critical $N_{\rm e}$ is 
1.6$\times 10^6$ cm$^{-3}$.  We calculate a similar critical
density for the [Ca \II] lines with the help of rates
calculated by Burgess et al. (1995), and a lifetime
of the $\rm ^2D$ term given by NIST (Ralchenko et al.
2010).

We see no evidence of the [S \II-2F]
pair at $\lambda\lambda$6717 and 6731 \AA, though they have
been observed in Herbig Ae/Be stars (Corcoran \& Ray 1997).
For this pair,
Draine gives critical densities of $10^3$--$10^4$ cm$^{-3}$.  We 
conclude the forbidden lines we do see
arise in a region where the electron density is 
between $10^4$ and $10^{6-7}$ cm$^{-3}$.

\begin{figure}
\includegraphics[width=55mm,height=83mm,angle=270]{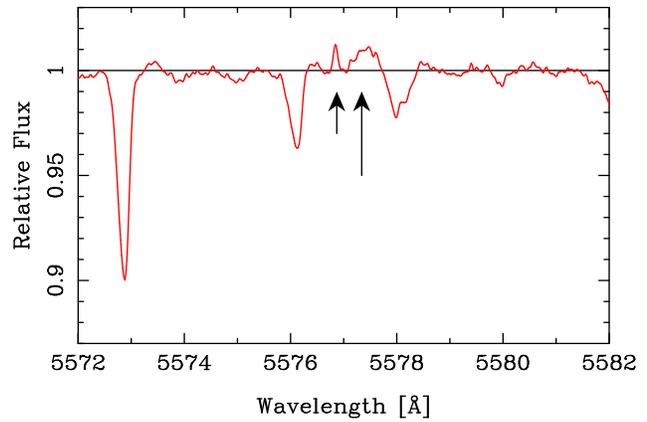}
 \caption{ The auroral ($\rm ^1D_2$\,--\,$ ^1\rm S_0$)
[O \I] transition $\lambda$5577 \AAß (gray, red in online version).
The long arrow points to 5577.34~\AA, the NIST wavelength.
The maximum, and center of gravity of the line shifted 
slightly to the red, as are the nebular lines (Fig.~\ref{fig:6300}).
The shorter arrow points to the sharp, blue-shifted feature
with the same displacement as 
the sharp component of [O \I] $\lambda\lambda$6300 and 6363 \AA.}
\label{fig:5577d}
\end{figure}

When the nebular as well as auroral transitions of [O \I]
are available, the ratio may allow one to determine values
of $T_{\rm e}$ and $N_{\rm e}$ compatible with the observation.  The average
of three measurements on $\lambda$5577 gives $W = 0.0042$\,\AA,
which with $W = 0.056\,$\AA\, for $\lambda$6300 yields a ratio
of 13.3.  If we assume the excited levels of O \I arise from
electron excitation, we may interpolate
in the plot of Gorti et al. (2011) to
find acceptable the values given in Table~\ref{tab:Gorti}.  
With the electron density constraint given above, we find 
temperatures in the range 7500 to 10000 K for the volume where
the forbidden lines are formed.

An alternate interpretation of [O \I] emission in Herbig Ae/Be
systems is discussed by Acke, van den Ancker \& Dullemond (2005).
In their model, the excited O \I levels arise primarily from
the photodissociation of the OH molecule.

%%\begin{center}
\begin{table}
\caption{Values of $T_{\rm e}$ (in K) and $N_{\rm e}$ (in cm$^{-3}$) compatible with the
observed $\lambda$6300/5577-ratio = 13.3.}
\label{tab:Gorti}
%\begin{center}
\tabcolsep=15pt
\begin{tabular}{r  c} \hline\noalign{\smallskip}
$T_{\rm e}$ ~~ & $\log N_{\rm e}$ \\[1.5pt]  \hline\noalign{\smallskip}
5000     & 8  \\
7500     & 7  \\
10000    & 6.5 \\
12000    & 6.2 \\[1.5pt] \hline
\end{tabular}
%\end{center}
\end{table}
%%\end{center}
\newpage
\acknowledgements

We are grateful for the availability of the ESO archive.
This research has made use of the SIMBAD data base, operated
at CDS, Strasbourg, France.  Our calculations have made extensive
use of the VALD atomic data base (Kupka et al. 1999), as well
as the facilities provided by NIST (Ralchenko et al. 2010).
CRC thanks colleagues at Michigan for many helpful suggestions.
Jes\'{u}s Hern\'{a}ndez suggested that we examine the forbidden
lines.

%\newpage%%%%%%%%%%%%%%%%%%%%%%%%%%%%%%%%%%%%%%%%%%%%%%%%%%%%%%

\end{document}